# A Proposed Experimental Test of Quantum Theory using the Techniques of Atom Optics

Peter J. Riggs

*Department of Quantum Science, Australian National University, Canberra, ACT 2601, Australia*

It may be possible to empirically discriminate between the predictions of Orthodox Quantum Theory and the deBroglie-Bohm Theory of Quantum Mechanics. An experiment using the measurement methods of Atom Optics is suggested in which an atom trap having evanescent light 'mirrors' in front of each of its walls is used to determine whether trapped atoms are in motion inside the trap. An absence of detected motion would be contrary to the prediction of Orthodox Quantum Theory and would support the deBroglie-Bohm Theory.

## 1. Introduction

The interpretation of the quantum mechanical formalism has a controversial history dating back to the theory's initial formulation and remains unresolved to this day. In Orthodox Quantum Theory (OQT), no physical reality is attributed to matter waves (as described by wavefunctions) [1–3]. Yet, convincing evidence for the objective existence of matter waves has been steadily mounting over the last two decades (e.g. [4–9]). In contrast, the deBroglie-Bohm (deBB) Theory of Quantum Mechanics (also known as the Causal Theory of Quantum Mechanics) postulates the existence of matter waves whilst making the same *statistical* predictions as does OQT [10, 11]. Given the accumulating evidence in favour of the reality of matter waves, it would be a highly significant breakthrough in quantum physics if an experimental test could be conducted that discriminates between OQT and deBB Theory. A proposal is presented here for such an experimental test employing Atom Optics. Similar to the case of the testing of Bell's Theorem, the conduct of such an experiment has had to wait until a sufficient level of advancement in measurement science has been reached – this has now been achieved (e.g. see: [12–19]).

## 2. Different Predictions

In order to devise an experiment that would empirically discriminate these two theories, there must be at least one situation where they make different predictions. Contrary to a common belief, it is not correct that deBB Theory produces exactly the same theoretical predictions to OQT in every conceivable circumstance. The 'infinite' potential well is one case where predictions differ in the two theories. Consider a neutral, spinless particle trapped in a rectangular box of side lengths $L_x$, $L_y$, $L_z$, with zero potential inside and 'infinite' potential outside (i.e. a three-dimensional 'infinite' well). Taking one corner of this well as the origin of a Cartesian coordinate system we find that a particle of mass *m* inside the well has the following stationary state wavefunction:

$$\Psi_{n_x n_y n_z} = (8/L_x L_y L_z)^{1/2} \sin(n_x \pi x/L_x) \sin(n_y \pi y/L_y) \sin(n_z \pi z/L_z) \exp(-iEt/\hbar) \quad \ldots (1)$$

where $n_x, n_y, n_z = 1, 2, 3, \ldots \infty$, $E = E_{n_x n_y n_z} = [(n_x^2/L_x^2) + (n_y^2/L_y^2) + (n_z^2/L_z^2)](\pi^2 \hbar^2/2m)$

is the particle's energy, t is time, and the other symbols have their usual meanings.

According to OQT, the particle inside the well is in motion as the ground state (kinetic) energy is non-zero. If the particle were not in motion then its momentum would be zero, contrary to the Uncertainty Principle (as understood in OQT) [20–23]. Since the particle in the well is *not* in an eigenstate of momentum [24], OQT requires measurements of the particle's momentum to result in values that occur with probabilities given by applying the Born Statistical Postulate. Then the actual momentum values found on measurement will occur with a frequency that closely approaches the probability P(**p**) of finding the particle with its momentum in a given range. This probability is calculated by integrating the momentum probability density $|\phi(\mathbf{p})|^2$ over a specified range in momentum space, i.e. P(**p**) = $\iiint |\phi(\mathbf{p})|^2 \, dp_x \, dp_y \, dp_z$ where $p_x$, $p_y$, $p_z$ are the Cartesian components of the momentum **p**, and $\sqrt{h} \, \phi(\mathbf{p})$ is the Fourier integral transform of $\Psi$ with $h$ being Planck's constant. The deBB Theory allows for the possibility where quantum equilibrium has not been established, i.e. where the Born Statistical Postulate (and therefore the standard quantum probability density) does *not* hold [25, 26].

In the three-dimensional well described above, the momentum probability density is [27]:

$$|\phi(\mathbf{p})|^2 = \{(2n_x^2 \pi L_x/\hbar)[1 - (-1)^{n_x} \cos(p_x L_x/\hbar)] / [(n_x \pi)^2 - (p_x L_x/\hbar)^2]^2\} \times$$
$$\{(2n_y^2 \pi L_y/\hbar)[1 - (-1)^{n_y} \cos(p_y L_y/\hbar)] / [(n_y \pi)^2 - (p_y L_y/\hbar)^2]^2\} \times$$
$$\{(2n_z^2 \pi L_z/\hbar)[1 - (-1)^{n_z} \cos(p_z L_z/\hbar)] / [(n_z \pi)^2 - (p_z L_z/\hbar)^2]^2\} \quad \ldots (2)$$

OQT predicts that various values of momentum will be measured with probabilities calculated by integrating Eq. (2) but the particle cannot be found to have zero momentum.

In deBB Theory, the configuration wavefunction of a quantum system $\Psi$ provides a mathematical description of its (physically real) matter wave. The wavefunction is expressed in polar form: $\Psi = R \exp(iS/\hbar)$, which is a natural expression representing the amplitude and phase of the matter wave, where R and S are real-valued functions of the space and time coordinates. The momentum **p** of a single particle is given by:

$$\mathbf{p} = \nabla S \ldots (3)$$

for a zero spin quantum particle [28, 29]. Inside the well, the particle is at rest because all of the particle's energy has become stored in the standing matter wave [30, 31]. The particle being at rest can be seen by applying Eq. (3) to the S function in Eq. (1), then $\mathbf{p} = \nabla S = \nabla(-Et) = 0$, i.e. the particle has zero momentum for all values of $n_x$, $n_y$, and $n_z$. This result can also hold when the well contains many particles. Since deBB Theory makes a different prediction to OQT for the 'infinite' potential well, the possibility of conducting an empirical test is opened up.

### 3. Reflection of Atoms from Evanescent Wave 'Mirrors'

This section will briefly review the relevant aspects of atom optics. When linearly polarized laser light is incident (at greater than the critical angle) on an interface between transparent mediums of higher to lower refractive indices, the laser beam undergoes total internal reflection.

An evanescent light wave is then generated on the lower refractive index side of the interface surface [32], which decays exponentially with distance from the surface. The potential of the electric field of this evanescent wave has the form $U_{ev}(z) = U_o \exp(-2\kappa z)$, where z is the direction perpendicular to the surface, $U_o$ is the value of the potential at the surface, and $(1/\kappa)$ is its decay length. In the case of a glass-vacuum interface, $\kappa = (2\pi/\lambda)(n^2 \sin^2\theta - 1)^{1/2}$ with $\lambda$ being the incident laser wavelength, $n$ is the index of refraction of the glass, and $\theta$ is the angle of incidence [33, 34] (as shown in Figure 1).

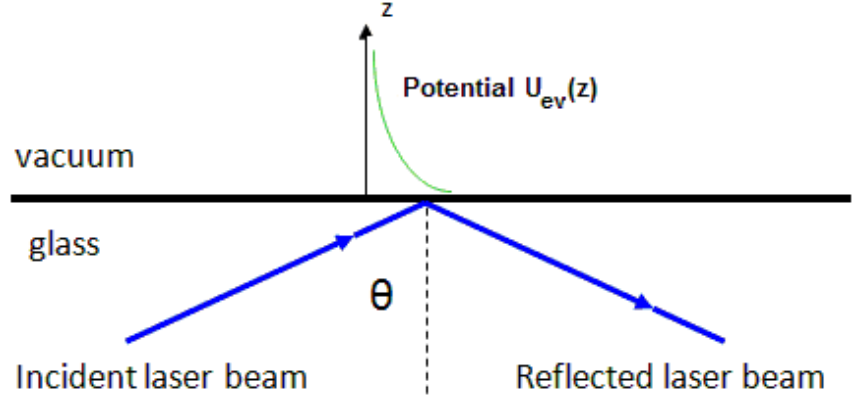

**Figure 1.** An evanescent wave is formed on the vacuum side of an interface surface when $\theta >$ critical angle.

The decay length depends on the intensity of the incident laser beam ($I_L$) and the angle $\theta$ [35, 36]. If the incident laser is transverse electric (TE) polarized then the generated evanescent wave will have intensity $I_{ev}$ given by [37, 38]:

$$I_{ev} = (4n \cos^2\theta) I_L / (n^2 - 1)$$

Suppose we have atoms that are two-level systems with resonance frequency $\omega_A$ (determined by an atom's energy level spacing), natural linewidth of atomic transition $\Gamma$, and saturation intensity $I_{sat}$ (= $2\pi^2 \hbar \Gamma c / 3\lambda^3$, where c is the speed of light in vacuum) [39]. Also assume a regime of coherent atom optics in which spontaneous emission is negligible (achieved when the detuning $\Delta = (\omega_L - \omega_A)$ is large, i.e. when $|\Delta| \gg \Gamma$) [40–42], where $\omega_L$ is the frequency of the incident laser beam. When $\Delta > 0$ (denoted 'blue detuning'), the interaction energy is positive and atoms of low energy entering the evanescent wave will be repelled without them reaching the interface surface [43–45]. The matter wave itself will extend to the surface. If an atom's speed is sufficiently low, it will be reflected elastically [46–49] and, for normal incidence, its (vector) momentum will become oppositely directed to what it was before reflection. Under these circumstances, the evanescent wave acts as an effective 'infinite' potential barrier to the atoms [50, 51]. The reflection of ultra-cold neutral atoms by this method is experimentally well established [52, 53]. Evanescent wave 'mirrors' may be used to construct a practical version of a three-dimensional 'infinite' potential well which will have desirable measurement advantages.

### 4. Proposed Test

In this section, an outline of an experimental proposal to test OQT is presented which was originally suggested some years ago [54] and is now feasible. The proposal is to make measurements inside a specially constructed atom trap (an effective three-dimensional 'infinite' potential well). The application of standard measurement methods inside particle traps, e.g. removing part of a trap's wall to facilitate measurement, introduce disturbances to the quantum system within [55–58]. Such disturbances limit the information on momentum that can be obtained to the distribution given by the Born Statistical Postulate. These restrictions have previously *ensured* that OQT and deBB Theory were empirically equivalent. The proposed experiment will avoid the limitations inherent in standard measurement methods by using atom optics techniques.

Information may be gained without introducing disturbances by reflecting atoms from evanescent wave 'mirrors' [59]. Following a suggestion by Cook & Hill and by Dowling & Gea-Banacloche, a suitable rectangular atom trap might be constructed [60, 61]. Laser light incident on the external walls of the trap totally internally reflects and generates evanescent waves on the inside of each of the walls. (In practice, the bottom of a super-polished prism would form each of the walls.)

Consider first the case where the trap contains many atoms. Imagine that a dilute 'cloud' of neutral, spinless atoms is placed in the centre of such a trap with the state preparation being an energy eigenstate. Prior to being placed in the trap, the atoms would need to be ultra-cooled to put them into their (preferably) lowest energy state, to have a deBroglie wavelength of the order of the trap's dimensions (so that wave effects dominate) [62], and to have a low enough speed to guarantee reflection by the evanescent light waves. When the atoms are in position, the lasers that generate each evanescent wave are activated. If the atoms are in motion once set inside the trap, they will be reflected when approaching close to any of its walls. The presence of atoms near to the trap's interior surfaces will result in small phase shifts in the reflected laser beams caused by slight changes in the refractive index of regions adjacent to the trap's walls [63, 64].

Aspect et al. calculated that if TE polarized lasers are used then the phase shift ($\varphi$) of a reflected laser beam due to atoms 'rebounding' is given by [65]:

$$\varphi = -\left(\frac{12}{\pi}\right)\frac{n\cos\theta}{(n^2-1)}\left(\frac{p^2}{M\hbar\Gamma}\right)\left(\frac{I_{\text{sat}}}{I_{\text{ev}}}\right)\left(\frac{\lambda^2}{\kappa}\right)\rho_{\text{in}} \dots (4)$$

where $p$ is the magnitude of the maximum momentum of the atoms, $M$ is the atomic mass, $\rho_{\text{in}}$ is the incident atomic density (i.e. the density when not close to the evanescent waves), and the other symbols are as previously defined. The measurement of a phase shift in any of the reflected laser beams constitutes, in the first instance, a measurement of the atomic density of the 'cloud' of atoms being reflected from the relevant trap wall [66]. Importantly, it is a quantum non-demolition measurement [67, 68] and no extra energy is imparted to the atoms [69].

Phase shifts in the reflected laser beams could be established by interferring each beam with a reference laser [70, 71]. The detection of phase shifts in one or more of the reflected laser beams over the lifetime of the trap would indicate that the OQT prediction that the atoms are in motion, is verified. The absence of phase shifts would indicate that the deBB Theory prediction is correct. Aspect et al. have suggested that such measurements would be feasible [72] (see also the discussion on phase shift enhancement in Section 5).

The absence of any phase shifts is dependent on the standing matter wave being undisturbed after the atoms are placed in the well, as disturbances to the stationary state would result in the atoms being accelerated [73]. However, determining whether the system has been disturbed or not may not be practical for a 'cloud' of atoms. The possibility that a disturbance to the stationary state of a many-atom system may be unavoidable in practice suggests that the experiment would be better conducted with just a single atom if measurements can be made sufficiently sensitive.

The use of only a single atom would avoid some complications, such as random collisions or the natural tendency for the atomic momentum distribution to move over time to $|\phi(\mathbf{p})|^2$ (i.e. moving to quantum equilibrium) [74]. An isolated, trapped, one-particle quantum system is achievable in practice [75–81]. Discussion of the conditions for detection of a single atom may be found in Courtois *et al.* (1995) [82]. The absence of any phase shift would indicate a lack of detection of the atom in motion, contrary to the OQT prediction.

The single atom case also offers the possibility of making individual momentum measurements which would permit a determination of whether the system is undisturbed or not. Since Eq. (4) shows that the phase shift depends on the atomic momentum, measurement of the phase shift would allow the momentum to be determined from the experimental data. If the OQT prediction is correct, then for multiple rebounds of a single atom, successive measurements of the phase shift in the reflected laser light should yield values of momentum with probabilities given by integrating Eq. (2) [83]. If measurements of the phase shift consistently showed values of momentum but with probabilities that differ from those predicted by OQT, this would indicate that a disturbance to the standing matter wave proved unavoidable in practice and the atom was thereby accelerated. However, these values of momenta with their different probabilities would also be acceptable as evidence against OQT and in favour of the deBB Theory.

## 5. Experimental Issues

The proposed experiment would be technically challenging. The use of neutral atoms is necessitated by the requirements of having spinless objects that do not interact with each other. Below are briefly mentioned a few of the practical experimental issues:

(i)     Effect of Gravity

The most obvious problem is the presence of the gravitational force. Although gravity will not affect the atom's motion along the length of an atom trap, it will affect vertical motion. This would result in a non-standing matter wave. Unless some means is devised to incorporate the effects of gravity without affecting the stationary state or otherwise prejudicing the measurements made, this experiment would need to be conducted in a *free-fall* environment.

(ii)    Vibrations of the Trap Walls

Vibrations of the walls of the atom trap would likely affect the standing matter wave and might possibly heat the atoms if they approach too near to the walls. Vibrations may be avoided by acoustically, mechanically, and thermally isolating the apparatus and ultra-cooling it. The apparatus should also be shielded from external light or other radiation.

(iii)   The van der Waal Interaction

The van der Waal interaction arises when atoms closely approach a surface (< 100 nm). This attractive potential falls off with the cube of the distance from the surface and reduces the height of the potential barrier in front of each wall [84]. The van der Waal interaction would need to be incorporated into a resultant potential [85, 86] which is the sum of the evanescent wave and van der Waal potentials. The resultant potential is still repulsive for distances at which slow atoms penetrate into the evanescent wave [87].

(iv)   Semiclassical Approach

Is a treatment where the atom is treated quantum mechanically but the evanescent wave is treated classically going to be adequate? What is essential for the proposed test is that the expression for the phase shift $\varphi$ holds, in order to determine whether or not the atom(s) are in motion. Treating the evanescent wave classically would seem to be sufficient for this purpose [88, 89], provided the practical issues (i) – (iii) listed above are dealt with appropriately.

(v)   Enhancement of the Phase Shift

Considerable enhancement of the phase shift can be achieved by having two dielectric layers deposited on each of the prism faces that form the walls of the trap [90–93]. Such enhancement may prove crucial to making accurate phase shift measurements [94].

(vi)   Homodyne Detection

In order to reduce fluctuations, each of the incident laser beams might be split into two parts with one part reflected from a wall of the trap and the other used as a reference beam [95].

## 6. Summary

An outline of how to perform one type of test of Orthodox Quantum Theory using atom optics techniques has been provided. If this experiment is conducted as proposed then the results should either confirm or disprove the prediction made by Orthodox Quantum Theory of the motion of atoms inside a three-dimensional 'infinite' potential well. Failure to confirm this prediction would also be evidence in favour of the deBroglie-Bohm Theory of Quantum Mechanics. This experimental test has the potential to deliver results with important implications for the understanding of basic quantum processes. The challenge is for experimentalists to devise a practical version of this test.